\documentclass[fleqn,twoside]{article}
\usepackage{espcrc2}
\usepackage{epsfig}

\def\nubar{\overline{\nu}}

\def\Cer{\v{C}erenkov{}}
       
\title{Neutrino oscillation studies 
and the neutrino cross section}
\author{ Paolo Lipari  \\
\small   Dipartimento di Fisica, Universit\`a di Roma ``la Sapienza",\\
\small   and I.N.F.N., Sezione di Roma 1, P. A. Moro 2, I-00185 Roma, Italy}

\begin{document}

\begin{abstract}
The present uncertainties in the knowledge of  the neutrino
cross sections   for $E_\nu \sim 1$~GeV, that is in the  energy range
most important for atmospheric  and long baseline accelerator
neutrinos, are large. These uncertainties  do not
play a  significant  role  in the interpretation of existing data,
however    they  could become a limiting factor
in future studies that aim at a complete and accurate determination
of  the neutrino oscillation parameters. 
New data  and theoretical understanding 
on  nuclear effects and on the electromagnetic  structure functions 
at low $Q^2$ and in the resonance production region
are available, and   can  be valuable 
in reducing the present systematic uncertainties.
The collaboration of physicists  working in different 
subfields  will be important to  obtain  the most 
from this available information.
It is  now  also  possible,  with the facilities
developed for long baseline beams,  to produce
high intensity and well controlled $\nu$--beams   to measure 
the neutrino  interaction properties with much better precision
that  what was done  in the past. 
Several projects and ideas  to fully exploit these possibilities 
are under active investigation.
These topics  have been the object of the  first
$\nu$--interaction (NUINT01)  workshop.
\end{abstract}

\maketitle

\section{Introduction}
Neutrino  physics is living in a ``golden era''.  Experiments 
with atmospheric  and solar neutrinos  have  given strong evidence
for the existence of  flavor oscillations,
or (less likely) some   other form  of `New Physics'  beyond
the Standard Model \cite{itow}.
The study of the $\nu$  flavor transitions  offer the 
possibility to obtain information about the neutrino masses and  mixing,
and the knowledge about these quantities
hopefully represents a fundamental window  on  the physics of the unification.
Theorists  are  busy  trying to  make sense  of the unexpected
results  that   have been obtained so far,
%Some  natural important  questions for example are:
%why  two  neutrino mixing angles ($\theta_{12}$ and $\theta_{23}$)
%are   large when all  quark mixing angles are small?
%Why is $\theta_{23}$ so close to its `maximal value' of  $45^\circ$ ?
%Why is  the  third mixing angle ($\theta_{13}$)  small? 
%What  sets the scale  of the $\nu$ squared mass differences?
on the other  hand  new experimental  studies 
are planned to measure 
 with greater  precision
the properties  of the   $\nu$ flavor transitions.
These   future studies,  are designed to  determine if the
$\nu$ flavor  evolution   is  completely described  by the 
3-flavor mixing  model, or if  a more complex dynamics
(transitions to additional sterile states, 
neutrino decay, flavor  changing neutral currents)  is  necessary,
and  should measure the entire set of parameters
(2 squared mass differences, 3 mixing angles and one CP--violating phase)
that determine the  $\nu$ flavor evolution in space--time
in the  3--flavor scenario.
\begin{figure}[hbt]
\centerline{\psfig{figure=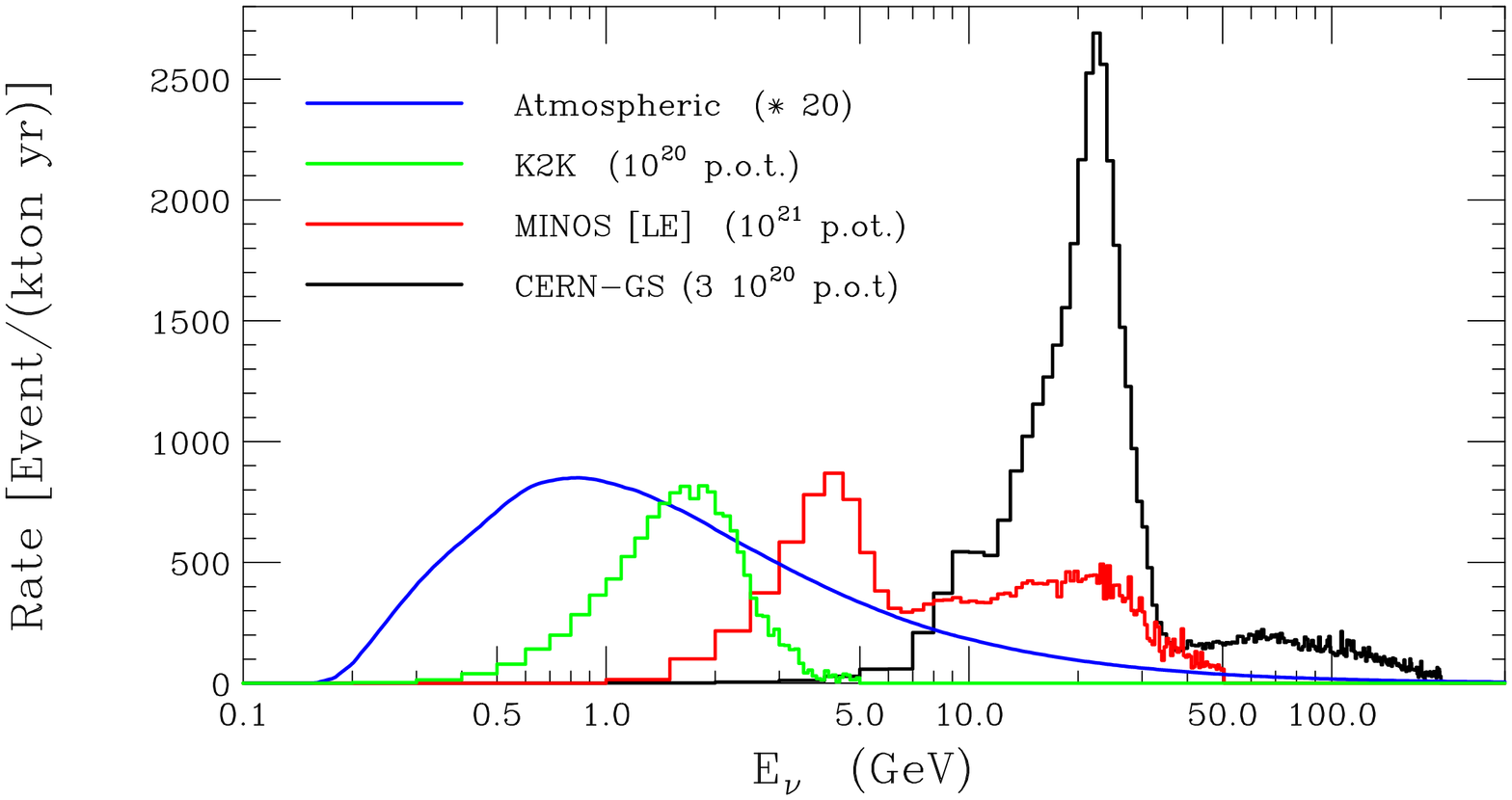,angle=90,height=4.2cm}}
\vspace {0.3 cm}
\centerline{\psfig{figure=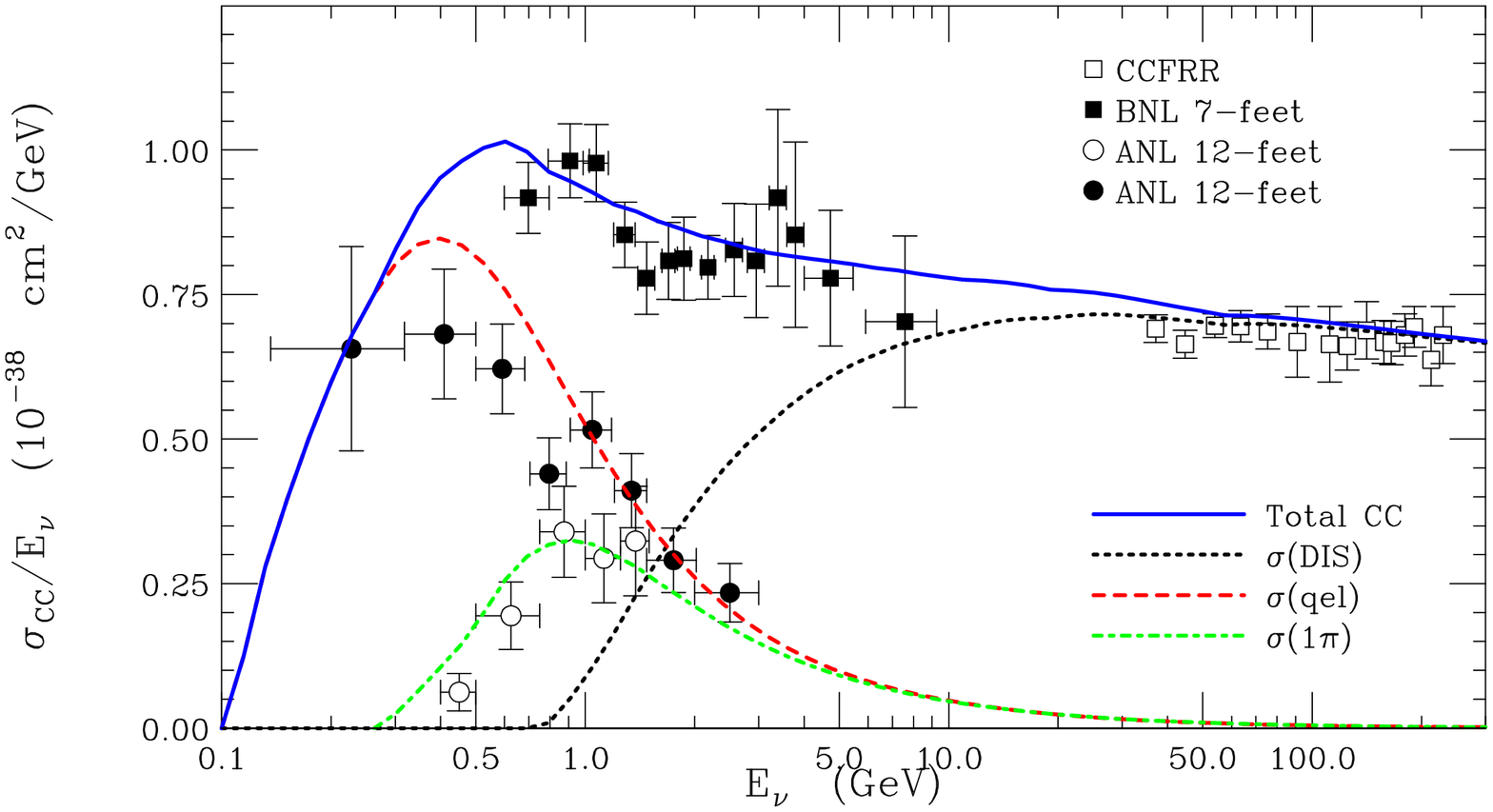,angle=90,height=4.2cm}}
\caption{\small \label{fig:spectrum}
Top panel: Energy distribution of the event rates ($dN_\mu/d\ln E_\nu$)
for atmospheric  neutrinos  and for  
 operating (K2K) or planned long baseline
neutrino experiments [Fermilab to Minos (LE) and CERN
to Gran Sasso].
Bottom panel: charged current $\nu_\mu$ cross section
calculation \protect\cite{LLS} compared with data.
Note the importance of quasi--elastic and single pion
production.
}
\end{figure}

The discovery of  $\nu$ oscillations has been performed
with natural (solar and atmospheric)  neutrino sources, however most of the
future  studies will be performed  with 
artificial neutrinos.
In particular, long baseline (LBL) accelerator  neutrino beams  will play 
a fundamental role. 
The first   of these projects,   the  KEK to Kamioka (K2K)
$\nu$--beam, with $\langle E_\nu \rangle \simeq 1.8$~GeV  and
$L \simeq 250$~Km,  has already collected
data, that support the existence of  the $\nu_\mu\to \nu_\tau$ transitions
as indicated by the atmospheric  neutrino data
\cite{itow,K2K}.  Other projects are  already
funded and in construction (Fermilab to Minos \cite{minos,para}, 
and CERN to Gran Sasso \cite{cngs,komatsu}, both
 with $L \simeq 730$~Km),  or in the design phase
 (the JHF project \cite{JHF,obayashi}).
Several `next generation'
 projects are also in discussion, involving ``super-beams'',
(very intense $\nu$--beams  produced with the traditional
technique: $p + {\rm Target} \to 
\pi^\pm, K^\pm, \ldots \to \nu_\mu (\nubar_\mu)$),
or  neutrino  factories (muon storage rings  that produce
extraordinarily well  controlled  neutrino beams  from 
muon decay: $\mu^+ \to \nubar_\mu + \nu_e + e^+)$.

The $\nu$~energy  (and pathlength) in the LBL projects 
is determined by    their scientific  goals and the existing
technical and experimental  constraints. 
The range of energy $E_\nu  \sim (0.2$--10)~GeV  
will have crucial importance. In fact 
the difference in quantum mechanical phase
developed by  two mass eigenstates  after propagation for a distance $L$
is:
\begin{equation}
\Delta \varphi_{jk} = 
(E_k - E_j)\, L \simeq 
{(\Delta m^2_{jk} \,L) ~/~  (2 E_\nu)}
\end{equation}
where  $\Delta m^2_{jk} = m_k^2 - m_j^2$
is  the $\nu$ squared mass difference. The probability for
flavor transitions  becomes   important  when at least one
quantum phase difference
is of order unity, and this determines the combinations
of $L$ and $E_\nu$ that are needed  to study  the flavor transitions
once the order of magnitude of  the $\Delta m^2$ are known.
The largest squared mass difference, measured in atmospheric  $\nu$ 
experiment: $|\Delta m^2_{\rm atm}| \equiv |\Delta m^2_{23}|$   is of order
$3 \times 10^{-3}$~eV$^2$.
A   quantum  mechanical phase difference  $\Delta \varphi_{23} = \pi$
corresponds to  the energy
$E_\nu  = {\Delta m^2_{23} \, L/(2 \pi)}$ or numerically:
\begin{equation}
E_\nu  = 0.60 \;{\rm GeV}
\left[ {\Delta m^2_{23} \over 3 \times 10^{-3}\;{\rm eV}^2} \right]
\left[ {L  \over 250~\;{\rm Km} } \right]
\end{equation}
At this energy one expects a maximum of the
$\nu_\mu \to \nu_\tau$ oscillation probability for a distance
$L$ between  $\nu$ source and detector.
Note that also the evidence for oscillation for atmospheric  neutrinos
has  been  obtained  with neutrinos  of energy 
$E_\nu \sim 1$~GeV.
The top panel of fig.~1 shows the energy spectra of  different
long baseline projects, and   for atmospheric  neutrinos\footnote{The
CERN Gran Sasso project is designed to study the  appearance
of $\tau$ neutrinos, with the 
process $\nu_\mu \to \nu_\tau \to \tau^-$. Since the threshold
for the  $\tau$ production is  $\sim 3.5$~GeV, the  beam is  designed to
have a larger average $\nu$ energy.}.

The flavor  composition, and energy spectrum 
of a   $\nu$ beam  in a detector
(located at a distance $L$ from the source)  is determined
from the observation of neutrino  interactions, 
and clearly  uncertainties in  the knowledge of the  neutrino
cross sections are a limiting  factors  in the sensitivity.
There are  uncertainties
about  the absolute value of  the cross sections,
the ratio $\sigma_{\rm nc}/\sigma_{\rm cc}$ 
between the neutral and charged current interaction rates, the
average  multiplicity in the final state, the  energy and  angular
distribution of final state lepton (and  hadronic  particles).
Depending of the detector   type  and the scientific  question
asked, these uncertainties  can have a different importance.

The NUINT01 workshop\footnote{See the web page
{\tt neutrino.kek.jp/nuint01/}}
has  been  dedicated to a discussion  of the problem  of the neutrino
interaction properties,  attempting to 
estimate: (i) the size of the  existing  uncertainties, (ii) their  probable
impact on studies of the fundamental  $\nu$   properties,
and (iii) possible methods to reduce these  uncertainties.
A list of   relevant questions  is:
\begin{itemize}
\item[(A)]  How well  have we measured the $\nu$  cross sections
and    interaction properties?

\item[(B)] How good is our theoretical	
understanding (and  computational  capabilities)
of the  $\nu$  cross sections  and interaction properties? 

\item[(C)] How well  do we {\em need} to  know $\sigma_\nu$ 
to  determine  with the desired accuracy the $\nu$ 
 flavor evolution and mass matrix?

\item[(D)] What experimental  programs   can/should be planned 
in order  to improve   our knowledge and understanding
of the  $\nu$  interaction properties?

\end{itemize}

\section{Measurements of $\sigma_\nu$}
The neutrino cross section can  be obtained from data,
from the formula:
\begin{equation}
\sigma_\nu (E_\nu)  = 
{N_{\rm int} (E_\nu) \over \varepsilon_{\rm det} \; \Phi_\nu (E_\nu) }.
\end{equation}
where $N_{\rm int}$ is the number of detected interactions,
$\Phi_\nu$ is the neutrino flux, and  $\varepsilon_{\rm det}$ is
detector efficiency.
This measurement has been performed in the energy region of interest
($E_\nu \sim 1$~Gev) by several  experiments\cite{ANL-qel}-\cite{Baker}.
The most accurate one  were  performed with bubble chambers 
filled with hydrogen or deuterium. 
A review of this data is given in \cite{sakuda}.
The  uncertainty of  $\sigma_\nu$  is dominated
by systematic errors in the  determination  of the $\nu$ flux.
For example   for the data of \cite{ANL-qel}
(the 12--foot hydrogen/deuterium
bubble  chamber at the Argonne National Laboratory),
the authors estimated
an   overall 15\% uncertainty for the  neutrino beam 
that was produced by   a 12.4 GeV proton beam  interacting on
a Beryllium target.
The systematic  uncertainty  was
estimated as as a convolution of different factors:
uncertainties on the fundamental   $\pi$ production cross section (5\%),
uncertainties of pion reabsorption in  the target  (5\%), proton  beam 
intensity (2\%), and smaller contributions.
Recent  critical analysis \cite{gaisser-honda}    of the  uncertainties
in  the $p$--nucleus  cross sections,   related to the calculations
of the atmospheric neutrino fluxes  have concluded  that the
uncertainties on the hadronic  cross sections are larger
(of order  15--20\%). To be conservative, it could be
safer to consider a systematic larger that what was quoted in the
original experiments.
The facilities  developed for the long--baseline  experiments
can provide $\nu$ beams significantly more intense  that the beams
used in the 70's.  In order to use these beams to obtain absolute values for
the cross sections it is essential to have good control of the
normalization and energy shape of the neutrino beam.

The K2K experiment has  already collected  in the near
detector  a very important sample of events
(see section 4), and 
a problem  of considerable interest is  to  determine   how well
the  $\nu$ beam is known.
The same question  clearly applies for the 
near detector  in the Fermilab to MINOS project.

In  principle it is possible to determine the    quasi--elastic
cross section  on (quasi)--free nucleons   
without  a precise  knowledge of the   $\nu$ flux, making use of
some theoretical input.
If the incident   $\nu$ energy is known,
from a measurement  of the energy and  direction of
 both final state particles in 
the scattering $\nu_\mu n \to \mu^- p$
one can determine both the neutrino energy  $E_\nu$ 
and the $Q^2$ of  the reaction\footnote{
The  $\nu$ reaction has to be performed
on a nuclear target (preferably deuterium)
and nuclear corrections have to be considered.}
It is therefore possible to
  determine the {\em shape}
of the  differential cross section $d\sigma/dQ^2 (Q^2)$,
that to a good  approximation can be considered as a function
of a  single parameter the axial mass $M_A$  (see next section),
while the absolute normalization, that is the value of the cross
section at $Q^2 \to 0$  can be inferred from $\beta$ decay  measurements.
This program   has  been performed in
\cite{BNL-7ft1}
(data from the  7--foot  hydrogen/deuterium  bubble  chamber 
at the Brookaven National Laboratory) 
with the result
$M_A = 1.07 \pm 0.06$~GeV, 
and in 
\cite{ANL-qel}
(ANL 12--ft bubble  chamber) with  the result
 $M_A = 1.01 \pm 0.09$~GeV. 
The total  cross section (and therefore the event rate) is
to a good approximation proportional to $M_A^2$. In \cite{ANL-qel}
the analysis of the event rate results in a lower estimate
$M_A = 0.75  \pm 0.12$. The poor agreement between the two
methods   suggests the existence of systematic effects.

Several  other experiments   have  measured $\nu$  interactions
in the energy range of interest using  nuclear targets. 
In the published analysis  of these works very often the
results are corrected for nuclear effects, and the data are the
expressed  as cross sections on free nucleons.
Since the nuclear effects  themselves are  difficult
to compute (as discussed in the following) and  the subject
of some controversy, some careful considerations are needed to use
these results.
For practical purposes it clearly would be very valuable to
have available also the results  for interactions on nuclear targets, 
before the nuclear corrections.

An example of this problem can be seen   analysing the results of the 
two  $\nu$ experiments \cite{ammosov,vovenko} operating 
with  the $\nu$--beam  at the Institute for High Energy Physics (IHEP)
at Serpukhov (Russia), that  were  presented at this workshop.
The  SKAT  bubble chamber \cite{ammosov}  has 
collected data on a set 
of low multiplicity  reactions,   namely
the final states 
$(\mu^- p)$, 
$(\mu^- p \pi^+)$, 
$(\mu^- p \pi^\circ)$, 
$(\mu^- n \pi^+)$
with an additional non--observed  `nuclear spectators' state.
Each one of the considered  final states  can  be  obtained 
as the result  of nuclear cascading processes, such as 
$\pi$  absorption ($\pi + A \to A^*$), $\pi$  production
($p p  \to p n \pi^+$, $p p \to p p \pi^\circ$, $\ldots)$;
and $\pi$ charge exchange
(for example  $\pi^+ n \leftrightarrow  \pi^\circ n$).
The experimentalists  have calculated the probabilities
for all these processes with montecarlo methods
to estimate the rate for exclusive  reactions
on free nucleons, 
however these corrections are probably the largest uncertainty
in the problem.
\begin{figure}[hbt]
\centerline{\psfig{figure=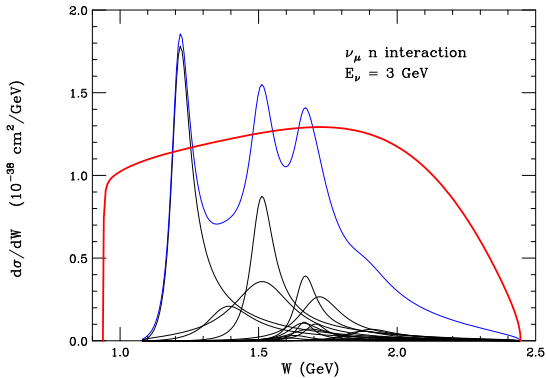,angle=90,height=5.0cm}}
\caption{\small \label{fig:sig_w}
Cross section for the  scattering $\nu_\mu + n \to \mu^- + X$.
The thick red line is   a calculation that  uses the DIS  formula
with the structure functions   given by the naive quark--parton model
and the PDF's of GRV--94 \protect\cite{GRV}.
 The other lines are the cross sections
for the exclusive  production of  a   resonance
($\nu_\mu + n \to \mu^- + R$)  and their sum,
according to the Rein and
Sehgal model \protect\cite{rein-sehgal}.
}
\end{figure}

\section{Cross section Formulae}
Only the  interactions of neutrinos with nucleons and  nuclei are relevant
for oscillation studies,  since the  ratio 
$\sigma_{\nu e}/\sigma_{\nu p(n)}$  is of order
$m_e/m_p$; 
the cross section can naturally be separated into
a charged current (CC) and a neutral current (NC) component:
$\sigma_{\nu_\alpha} = 
\sigma_{\nu_\alpha}^{\rm cc} + \sigma_{\nu_{\alpha}}^{\rm nc}$.
For low energy neutrinos  it is also natural to decompose
the cross section according to the multiplicity:
$$
\sigma_{\nu_\alpha}^{\rm cc} = 
\sigma_{\nu_\alpha} ({\rm qel}) + 
\sigma_{\nu_\alpha}^{\rm cc}(1\pi) + \sigma(2\pi) +\ldots
$$
The  correct calculation of  the quasi--elastic and
one pion components 
 has  great importance in the analysis
of existing data since  most of the data on 
atmospheric  neutrinos  has  been obtained with 
water \Cer{} detectors (Super--Kamiokande, and before IMB and
Kamiokande) that, because of relatively  poor pattern recognition
capabilities,  have  limited most of their analysis to the so called
single--ring events, where only one  particle is visible.
This selects quasi--elastic  scattering\footnote{Most final state
protons are below the \Cer{} threshold.}
with in addition events with one (or more)
charged pion(s) below the \Cer{}  threshold.
The  bottom panel  of fig.~1 shows a plot of $\sigma_{\nu_\mu}^{cc}$
separated  according to the pion multiplicity and compared
with some existing data.  The importance of the low
multiplicity channels is clearly displayed

\subsection{Quasi--elastic scattering}
The quasi--elastic cross section can be written in terms of 
nucleon form factors.
The most general matrix  element for a nucleon transition can be written
\cite{Llewellyn} in terms of six  form factors:
%\begin{equation}
$$
\begin{array}{l}
\langle p(p')|J^\alpha(0)|n(p) \rangle = \overline{u}_p(p')
 \\
~~~ \\
~\left[
\begin{array} {l} 
F_V(q^2)\gamma^\alpha+iF_M(q^2)\sigma^{\alpha\beta}\frac{q_\beta}{2M} \\
F_A(q^2)\gamma^\alpha\gamma_5+F_P(q^2)\frac{2M}{{m_\pi}^2}q^\alpha\gamma_5 \\
+i{F_A}^{\rm sc}(q^2)\sigma^{\alpha\beta}\frac{q_\beta}{2M}\gamma_5
+{F_V}^{\rm sc}(q^2)\frac{q^\alpha}{M}
\\ \end{array}
\right ]u_n(p) \\
\end{array}
$$
%\end{equation}
The hypothesis of CVC  allows  to relate the vector form factors $F_V$ and
$F_M$ to the electromagnetic  form factors of the proton and  neutron;
the  contribution of second class currents (that violate $G$--parity)
can be assumed to vanish, and one remains with two unknown
functions the axial form factor $F_A$ and the pseudoscalar form factor
$F_P$.  The  contribution of the
 pseudoscalar form  factor to the cross section  is proportional to
$m_\ell^2/M^2$, 
therefore this  term is important only for $\nu_\tau$,  while  the axial
form factor $F_A$  remains  undetermined.
The value of $F_A$ at $Q^2 =0$ can be related to the axial  coupling measured
in $\beta$-decay experiments 
($F_A(0) = g_A = 1.2670 \pm 0.0035$), 
but the $Q^2$  dependence  of $F_A$ has to be determined  experimentally.
An often used parametrization is a dipolar form:
\begin{equation}
F_A (Q^2) = g_A~\left [ 1 + { Q^2\over M_A^2 } \right ]^{-2}
\label{eq:fa}
\end{equation}
that depends  on the  single  parameter $M_A$.
Therefore in first approximation 
the problem of the determination of the quasi--elastic cross section
can be indentified with  the measurement for $MA$.
A review of the status of our knowledge of the nucleon form factors
was  given in \cite{singh} and \cite{thomas}.

\subsection{Inelastic channels}
The  kinematics  of a final state lepton is described 
by  two  variables,
that correspond to  its  energy  $E_\ell$ and angle with respect
to the initial neutrino $\theta_{\ell\nu}$. The   kinematical variables 
can  be  chosen as  relativistic  invariants  such as
the adimensional scaling  variables $x$ and $y$;  
of particular dynamical importance are the 
transfer momentum in the interaction: 
quantities:
\begin{equation}
Q^2 = -q^2 = -(p_\nu - p_\ell)^2,
\end{equation}
and the square of the hadronic mass in the final state:
\begin{equation}
W^2 = p_{\rm had}^2 = (p_i + q)^2.
\end{equation}
The  kinematically allowed range of $Q^2$  is the interval
between  $Q^2_{\rm min} \simeq 0$  and
$Q^2_{\rm max} \simeq 2 m_p \,E_\nu$  that
grows  linearly with energy. For  $Q^2_{\rm max} \ll M_W^2$
(that is for $E_\nu \ll 3$~TeV) the effects  of the
$W$ propagator are small and the cross section
 $\sigma_\nu$ grows  approximately
linearly with energy.
Similarly  the allowed  range of  the squared hadronic mass  $W^2$ is 
between  the values $m_p^2$  and 
 $W^2_{\rm max} - m_p^2 \simeq 2 m_p \, E_\nu$.

In general \cite{paschos,bodek}, the inclusive differential  neutrino cross section 
can be written  in  terms  of 5 
structure functions  $F_j(x,Q^2)$,  according to the general  formula:
%\begin{equation}
$$
\begin{array}{lcl}
{{\rm d}\sigma^{\nu,\nubar} \over {\rm d}x{\rm d}y}
& = &
     \frac{G_F^2 M_N E_\nu}{\pi} 
\; \left ( { M_W^2 \over M_W^2 + Q^2} \right )^2 \times  \\

& ~ &   \bigg \{ y\Big(xy + \frac{m_\ell^2}{2 E_\nu M_N}\Big)F_1 \\
& + & 
    \Big(1-y -\frac{M_Nxy}{2 E_\nu} - 
     \frac{m_\ell^2}{4 E_\nu^2}\Big) F_2  \\
&\pm & 
\Big(xy(1-\frac{y}{2})-y\frac{m_\ell^2}{4 M_N E_\nu}\Big) F_3 \\ 
& + & 
\frac{m_\ell^2}{2 M_N E_\nu}\cdot \Big [
\Big(x y 
+ \frac{m_\ell^2}{2 M_N E_\nu}\Big) F_4  - F_5  \Big ]
\bigg\}, 
\end{array}
$$
\begin{figure}[hbt]
\centerline{\psfig{figure=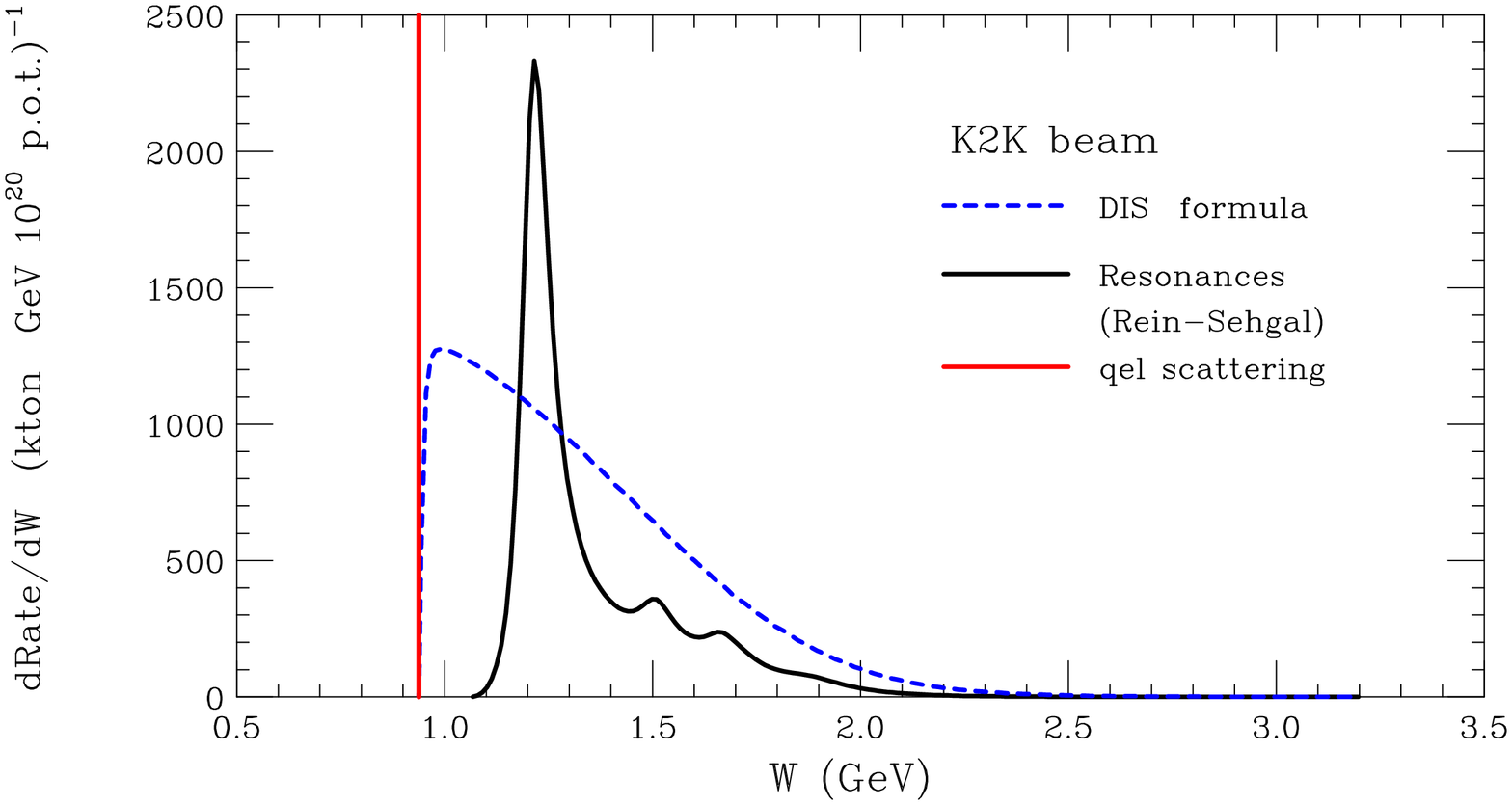,angle=90,height=4.2cm}}
\vspace {0.3 cm}
\centerline{\psfig{figure=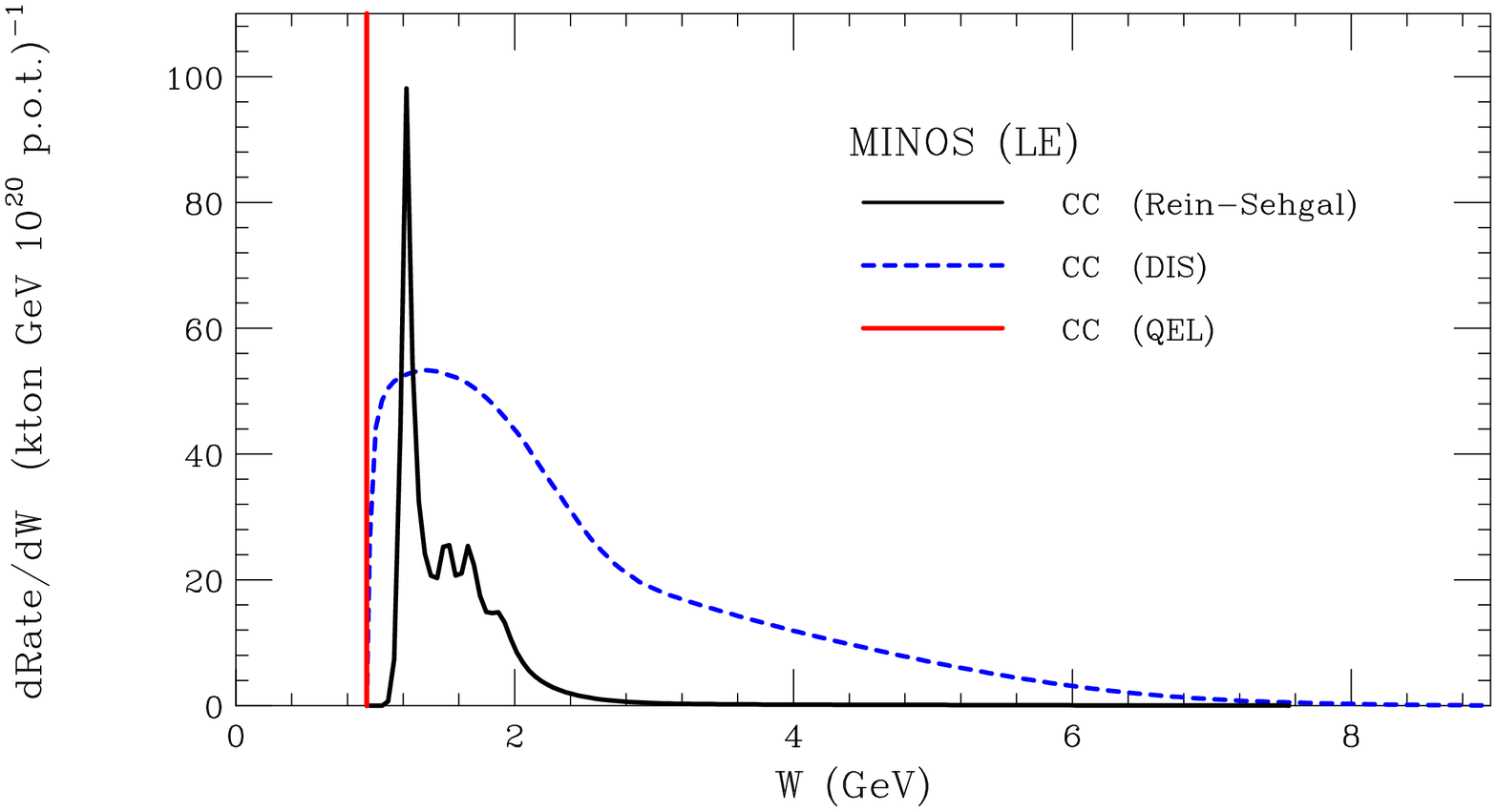,angle=90,height=4.2cm}}
\caption{\small \label{fig:rate}
Calculated distributions of  the differential cc event rates 
($dN_\mu/dW$)in the far detector
as a function of $W$ the hadronic mass in the final state.
The top panel is for the K2K experiment, the bottom panel for
the Fermilab to Minos project.
The vertical  line at $W = M_N$ represents  quasi elastic scattering,
the dashed line is calculated according to the DIS formula,
the solid line is  the contribution of the resonances according
to the Rein and Sehgal model \protect\cite{rein-sehgal}.
}
\end{figure}
%\end{equation}
where the two signs for the $F_3$ component 
correspond to $\nu$ and $\nubar$.
The scaling variables $x$ and $y$ are defined as
$y = (q\cdot p_i)/(p_\nu \cdot p_i)$ and
$x = Q^2/(2 M_N \,E_\nu y)$.
 Note that the structure functions
$F_4$ and  $F_5$  contribute terms  proportional to
$m_\ell^2/(2 M_N \,E_\nu)$ and are therefore of practical interest
only in the case of  $\nu_\tau$ cc interactions 
(where they constitute a source  of uncertainty).
It is  well known that in the Deep Inelastic Scattering (DIS) limit,
that corresponds to  $Q^2$ and $W^2$ large, the 
structure functions can be related to the
Parton Distribution Functions that can be well measured in different
experiments, and have a calculable $Q^2$ evolution. For example
in leading order  of an $\alpha_s$ expansion:
\begin{equation}
\begin{array}{lcl}
F_2^{\nu, {\rm cc}}  & = & 2 x \, F_1^{\nu, {\rm cc}}  =
2x \, (d + s + \overline{u} + \overline{c}) \\
F_3^{\nu, {\rm cc}}  & = & 2 \, (d + s - \overline{u} - \overline{c})
\\
\end{array}
\end{equation}
\begin{figure}[hbt]
\centerline{\psfig{figure=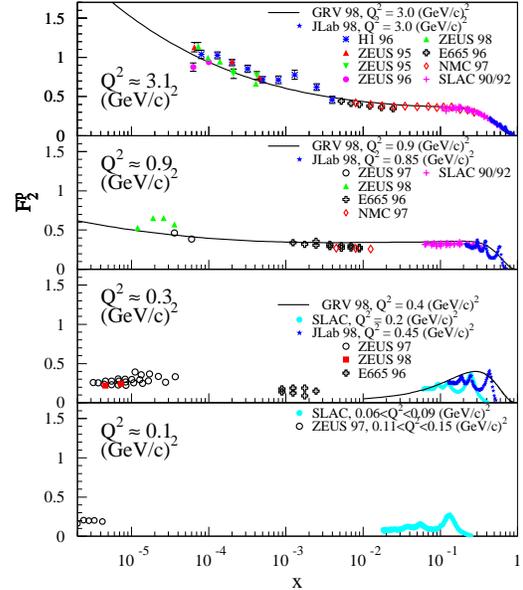,width=7.4cm}}
\caption{\small \label{fig:f2_a}
Structure function $F_2^{\rm e.m.}$ as 
a function of $x$ for four values of $Q^2$,
with a logarithmic $x$ scale. The symbols indicate various experiments.
The solid curves in the top two panels represent the
calculated distributions from the GRV collaboration \protect\cite{GRV},
evolved from $Q^2$ = 0.4 (GeV/c)$^2$. The solid curve in the third panel
represents the input distribution itself.
From \protect\cite{nicu1}.
}
\end{figure}
Therefore  in  the DIS region  the neutrino cross section is
well  understood and under good theoretical control.
However,  essentially the entire  phase space relevant for
atmospheric  neutrino contained events,  
is  well outside of the DIS region, and  this  remains true 
for the K2K, Minos to Fermilab and JHF projects.
An illustration of this  problem  is  shown in  fig.~2, where 
we plot the cross section  for the scattering
$\sigma_\nu + n \to \mu^- + X$ at an energy of 3~GeV.
At this  energy (that is actually rather  high for
atmospheric   $\nu$ and  most LBL experiments),
the range of possible $W$ is small  and extends in the region  
where hadronic  resonances are present. 
The naive use of the DIS formula  with the 
PDF's of Gluck Reya and Vogt \cite{GRV}  results in a smooth
curve, that predicts a non vanishing cross section
also in  the  interval $ m_p \le W \le m_p + m_\pi$.
Obviously  a realistic  calculation
must predict a more complex   structure, as it is  the case 
for the model of Rein and Sehgal \cite{rein-sehgal} whose results
are also  shown in the figure.
Figure~3  is another illustration of the importance of the resonance region 
in the K2K and Fermilab to MINOS projects.
The delta function  for $W = m_p$   has an area
of $\sim 42\%$ for K2K  and of order 11\% for MINOS (LE beam). 

\begin{figure}
\centerline{\psfig{figure=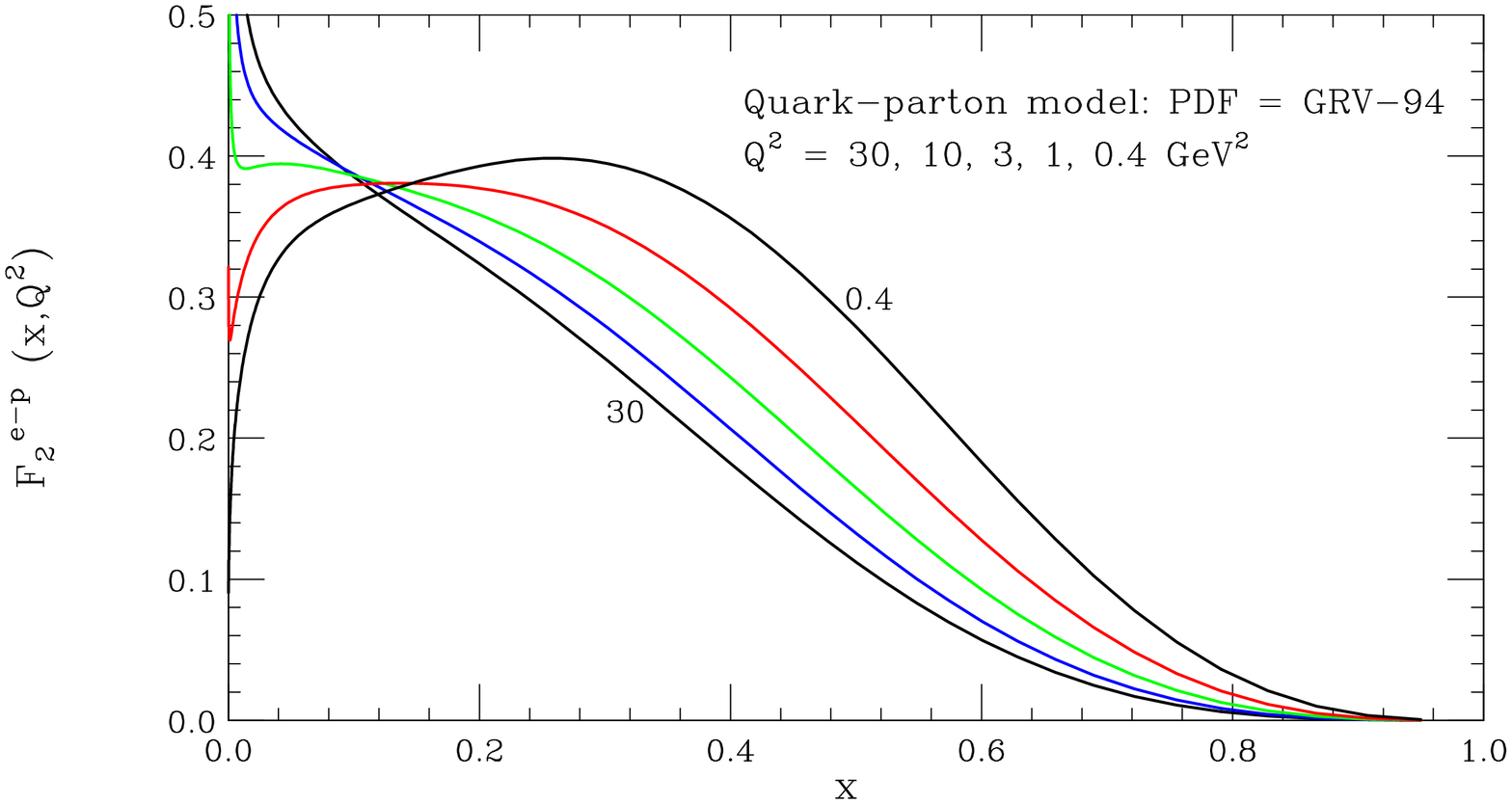,angle=90,width=7.4cm}}
\vspace {0.3 cm}
\centerline{\psfig{figure=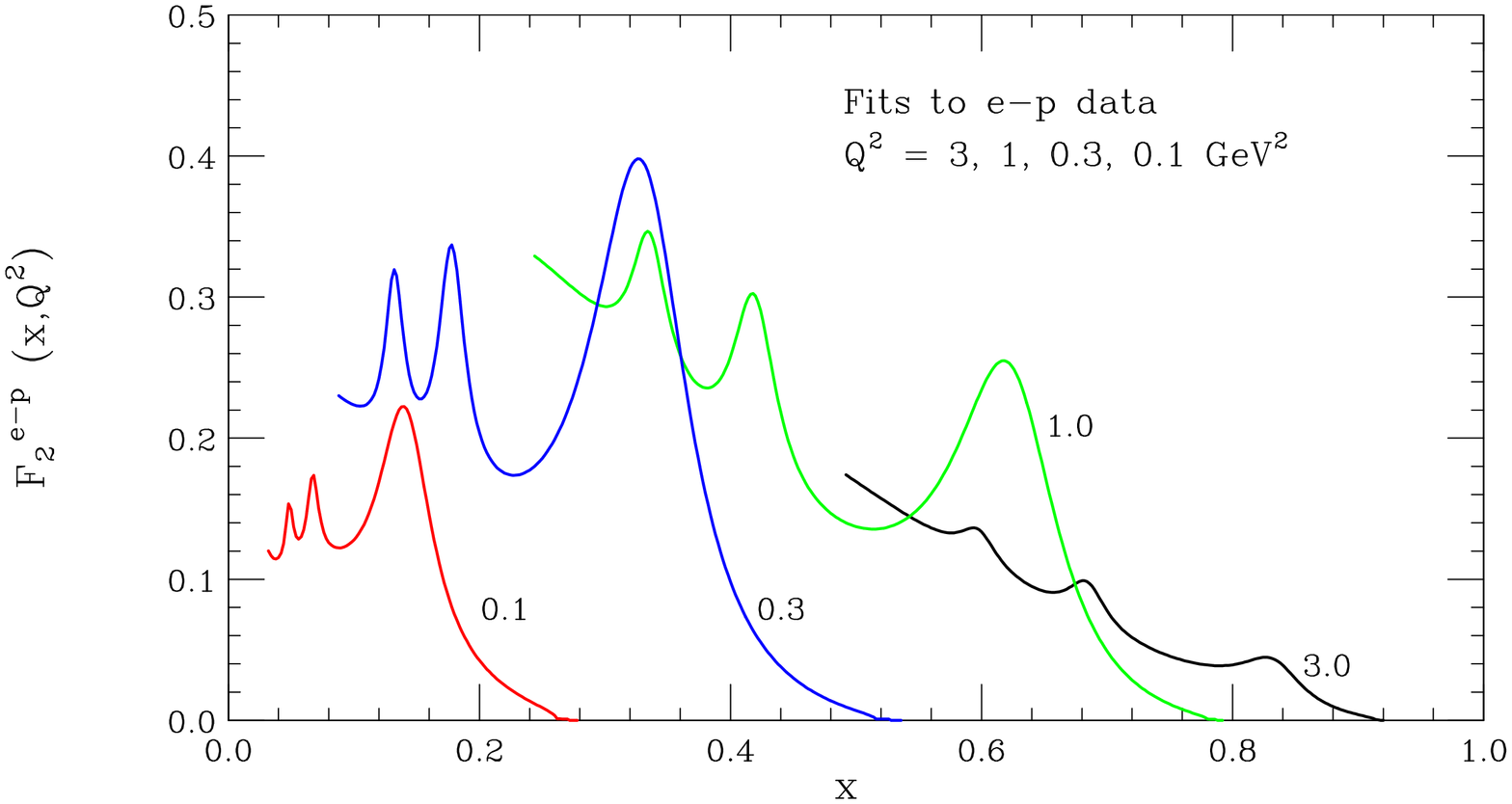,angle=90,width=7.4cm}}
\caption{\small \label{fig:f2_b}
Structure function $F_2^{\rm e.m.}(x, Q^2)$  for $e$--$p$ scattering.
The top panel  shows  what is obtained from the simple use of the naive
quark--parton model.  The bottom panel shows  fits to the
SLAC and JLab data.
}
\end{figure}

A central problem for the calculation of the $\nu$ cross section
is the correct description of the   low $Q^2$, low $W^2$ kinematical  range,
where perturbative QCD is not  valid.
Luckily, in recent times, a large  body of highly accurate data 
on electron--proton  scattering in this  kinematical  region
  has  become  available, mostly
from the Jefferson laboratory (see \cite{nicu1,wood,kinney,bodek}).
An example of this  data is shown in fig.~\ref{fig:f2_a}
where is plotted the 
electromagnetic structure function $F_2^{\rm e.m}$ that in
the DIS limit in leading order can  be expressed in terms of
PDF's as:
\begin{equation}
F_2^{\rm e.m.} = \sum_j e_j^2 \, x \, [ q_j(x) + \overline{q}_j]
\end{equation}
One can see that the production of resonances  plays a  very important role, 
and that three mass peaks are clearly visible at low $Q^2$.
Fig.~\ref{fig:f2_b}  also   illustrates the importance of   a
proper  treatment of the  relevant effects in the low $Q^2$,
region.
The physical  insight that can  be obtained  from the electron scattering
data  can  be applied also to neutrino interactions,
even if significant   theoretical  uncertainties remain.
A concept of   great importance in the    resonance  region is the 
notion of ``parton--hadron'' duality.  Already in 1971,
Bloom  and Gilman  \cite{bloom-gilman} observed  empirically,
 that the  scaling  structure  function  $F_2^{em} (x)$ 
that characterizes   electromagnetic scattering
in the   DIS (high $Q^2$ region), represents a good  average   over
the resonance bumbs  of the inclusive data  observed at much lower 
$Q^2$. The agreement can  be significantly improved  rescaling the
$x$  variable for   target mass correction  and using for
example the Nachtmann variable $\xi = 2x /(1 + \sqrt{1 + 4 M^2 x^2/Q^2})$.

\subsection{Nuclear effects}
\begin{figure}
\centerline{\psfig{figure=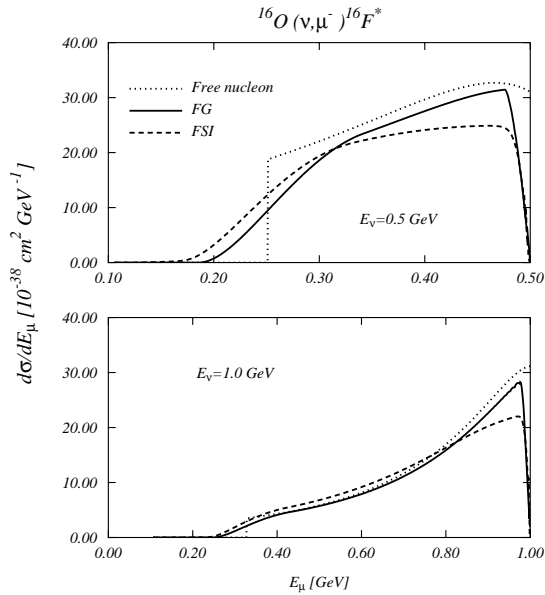,width=7.4cm}}
\caption{\small \label{fig:fermi}
  Cross section integrated on the muon emission angle as a function of
  the muon energy.
 The   full lines have been calculated with the Fermi gas model (FG label)
 and the   dashed lines  with  the model described in   \protect\cite{g.co}
that includes final  state interactions (FSI label).
(from \protect\cite{g.co}).
}
\end{figure}

Neutrino  detectors  are usually composed of 
heavy nuclei (the  most important  nuclear species are 
oxygen (SK), argon  (Icarus),
iron (Minos) and lead (Opera)),  and in the energy range  we are discussing
nuclear effects are important and have to be taken properly into account.
Schematically one can separate the nuclear effects into two  parts:
(i) the  $\nu$ interaction with a bound nucleon, and
(ii)  the  propagation of the final state  hadrons in  the nuclear matter,
even if  a clear distinction of these two effects is not  rigorously valid.
The  simplest and  most commonly used approximation
to describe  the interaction with a bound nucleon 
is the Fermi gas model.
In this model  the nucleus is  described  as an ensemble of 
nucleons  with a momentum distribution  $f_{p,n}(\vec{p}_i)$.
The  energy    that corresponbds to the 
3--momentum $\vec{p}_i$ is $E_i = \sqrt{p_i^2 + M^2} - \varepsilon_B$ where
the quantity $\varepsilon_{B}$ is the binding energy  that 
can be different for protons and  neutrons\footnote{In some
implementation of the model   $\varepsilon_B$ is also a function
of $\vec{p}_i$.}.
In the simplest version of the model \cite{smith-moniz},
the  initial nucleon momentum  distribution has a 
``zero temperature''  shape:
\begin{equation}
f(\vec{p}_i) = \Theta (p_{F} - |\vec{p}_i|)
\end{equation}
that  vanishes  when  $|\vec{p}_i|$ is   larger   than the 
Fermi momentum $p_F$, in  other models it has a more complex shape
that includes  higher momentum  tails.
Quasi--elastic scattering can be simply described
as an   elementary ($2 \to 2$) process such as
$\nu_\ell + n^* \to \ell^-  +  p^*$   where $n^*$ and $p^*$ are off--shell
initial final  states nucleons with 3--momenta 
$\vec{p}_i$ and $\vec{p}_i +\vec{q}$ ($q = p_\nu - p_\ell$ is the 
transfer 4--momentum)  and binding energies
$\varepsilon_i$ and $\varepsilon_f$.
This procedure allows  naturally to  
implement  Pauli--blocking,   forbidding  the scattering
into occupied  nucleon states. This can be done  including a factor
$[1 - f_p (\vec{p}_i + \vec{q})]$ in the calculation 
of the matrix element  for the scattering.
Some  natural questions, that have been  discussed at the workshop are:
(i) how good is the Fermi gas model  as  a description of
the nuclear effects;
(ii)  how large are the expected  corrections to the Fermi gas model;
(iii) what is the best (or most convenient)   theoretical  framework
to  describe  nuclear effects  beyond the  Fermi gas model.
It is clear  that conceptually the  Fermi--gas model is  a very simple
and non--realistic description of the scattering \cite{pandha,miller},
it is however not  entirely  clear   how large is  the  error  introduced by 
this  simplified treatment. 
Before, during  and after  the workshop,  it was  attempted to 
produce and compare different ``beyond the Fermi--gas model''  calculations,
(see  the critical discussion by R.~Seki  \cite{seki} in these Proceedings).
Hopefully a  generally  accepted  model  will soon emerge
from this work. An example  of the results  for one  of these more
sophisticated models (by G.~C\`o and  collaborators \cite{g.co}). is shown 
 in fig.~\ref{fig:fermi}. This model  predicts  
a suppression of the cross section by approximately 15\% with respect to
the Fermi gas model, and a softer spectrum for the final state lepton.

The  uncertainties  due to nuclear effects  become larger
when discussing   the resonance production processes and the propagation
of hadronic particles in the  nuclear matter.
It is important to notice  that because of pion reabsorption  processes
the calculation of pion--less final states must also include
contributions from resonance production.

\section{The K2K  near detector data}
The K2K experiment   has  collected  data with  the 
neutrino  beam   obtained   from  a  12 GeV protons beam
at the KEK accelerator.  The goal of the experiment is  to
study the existence of neutrino  flavor transitions  comparing
the event  rates at a  near detector, located
approximately 300~m from 
the  pion production target,
with event rate  at a far detector(Super--Kamiokande)  located
at a distance of 250~km.
The    event  rate  recorded at the far detector   has been 
measured to  be smaller that the rate  extrapolated from the 
near detector results, with a suppression that is consistent  with
the $\nu_\mu \leftrightarrow  \nu_\tau$ oscillation interpretation
of the atmospheric  neutrino data \cite{itow,K2K}.  

The  neutrino  interactions  collected at  the near detector
represent a very large sample with potentially extremely useful
information. The discussion of  this data has  been perhaps the
highlight  of the workshop.

The near detector for the K2K    experiment  is really
a combination of three detectors.
The main flux measurement is based on a one-kiloton water 
\Cer{} detector.
This detector 
uses the same photo-multipliers as Super--Kamiokande,
and has  essentially the same  design (except for scale) and the analysis
for the two detectors  can use  the same reconstruction algorithms.
A fine grained detector sits downstream of the  water detector.
It  is made as a  scintillating fiber tracker 
with water targets enclosed between 
layers of tracking material.
The next component of the near detector
 is a stack of iron plates interleaved with
drift tubes to serve as a muon range detector.
Both the fine--grained  and   muon range detectors
have transverse pairs of tracking planes so
tracks can be reconstructed in three dimensions.

The data  of the three  near detectors  has  been discussed in a set of
4 contributions
by Ishida, Walter, Mauger and Mine
\cite{ishida,walter,mauger,mine}.
The event sample collected up to now in the near detectors amount
to a  very  significat statistics:
27,000  events  in  the water detector, 8,300 in the fine grained detector
and 125,000 in the iron detector.
A comparison  of  the data with a Montecarlo simulation shows 
in general  good agreement, demonstrating that there are no  large
systematic errors in the simulation chain, 
however small but significant effects  are visible.

\begin{figure} [hbt]
\centerline{\psfig{figure=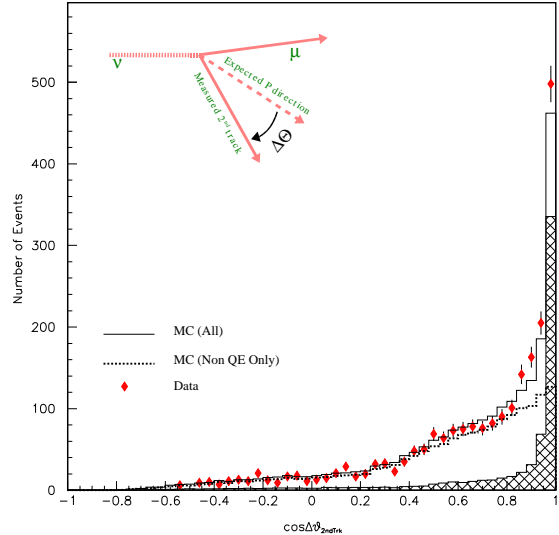,width=7.4cm}}
\caption{ \small 
\label{fig:finegrained}
Results from 2-track events in the  K2K fine-grained
near detector. The abscissa is the cosine of the angle between
         the second track reconstructed direction and the direction calculated
         for a proton based on the assumption that the other track was a muon 
         and 2-body kinematics as is shown in the inset cartoon.
The  fit of  this distribution
can give a solid measurement of the
non-quasi-elastic  contribution  in the data.
}
\end{figure}

The fine grained detector \cite{walter}
 as a whole is excellent for event classification
studies since it can distinguish quasi-elastic and inelastic 
events.  This is  performed  measuring  the   
momenta and directions of
the muon and the  proton in events of type:
$\nu_\mu + A \to \mu^- + p (+ \ldots)$.
In case of quasi--elastic   scattering on a free nucleon,
because of 4--momentum conservation, 
the  information obtained from the 
the measurements  is redundant, and   
the measurements  of $E_\mu$, $\theta_{\mu\nu}$ and  $E_p$   allow 
to  determine  the initial   neutrino energy  
 and  the final state proton direction.
For scattering on a bound nucleus,  since the   spectator
nucleons   absorb some 4--momentum, the   predicted values
represent  the central   values  of narrow distributions.
The contribution of  the inelastic channels  can be   separated
from the quasi--elastic  one because it  is much broader 
as  illustrated in  fig.~\ref{fig:finegrained}.

A method to 
 discriminate
between the hypothesis of
$\nu_\mu \to \nu_\tau$  
and 
$\nu_\mu \to \nu_{\rm ster}$  transitions is the 
study  of the neutral  current     event rate.
 This  event rate is suppressed 
in the case  of  transitions of $\nu_\mu$ into sterile neutrinos,
while it  remains  constant for  standard
$\nu_\mu \to \nu_\tau$ transitions.
The most important channel for the study of the neutral  current event rate 
is  the  reaction:
$\nu_x + A \to \nu_x + A + \pi^\circ$
that has  the lowest multiplicity and a clear experimental signature.
The problem is that   in order to predict the  event rate
for this  process one  needs to know with sufficient precision
the absolute value  of the cross section, or at least
the ratio  with respect to a measurable  charged  current process.
The K2K near detector is the ideal instrument to measure
this  cross section (or  a cross section ratio  such as
$\sigma_{\rm nc}(1\pi^\circ)/\sigma_{\rm qel}$.
The status of these measurements is described in  \cite{mauger}.
Approximately 5000 $\pi^\circ$ were reconstructed  in a 50~ton 
fiducial volume in the water \Cer{} near detector,
determining the ratio $(\pi^0/{\rm FC}_\mu$) with small statistical error
and an overall systematic error of order $\sim 9\%$.
This can be applied to the 414 reconstructed $\pi^\circ$ in the atmospheric
neutrino sample (1289 days) of SK to obtain a result  that
disfavors the oscillation into  sterile neutrinos.

It is well known that  atmospheric  neutrinos 
were initially considered mostly as an unpleasant background
for proton decay studies, because  for any  detector resolution,
there is a finite probability
that a $\nu$ interaction  will mimic
a proton decay   event.  It is  clearly  very important to 
estimate correctly this probability   for the interpretation of
 the existing data, and more important for the planning of future
experiments. This probability is clearly 
strongly dependent on the $\nu$ interaction properties.
The exposition of the 1~kton  (50~tons fiducial mass) water \Cer{} detector 
 to the  $\nu$  fluence
obtained from $3 \times 10^{19}$ protons  is   roughly equivalent
to  a $\simeq 1$~Mton~yr   exposition  to the atmospheric  $\nu$--flux
and therefore allows to study  experimentally 
 the importance of the atmospheric background
for $p$ decay  studies (see \cite{mine} for more discussion).

\section{How well do we {\em need} to know the $\nu$ cross sections?}
It is  necessary to estimate carefully the
possible impact  of 
systematic uncertainties in our knowledge of the neutrino
cross sections for  the interpretation of present and  future
neutrino data.

The effect of uncertatainties on $\sigma_\nu$ for the 
interpretation of the SK atmospheric neutrino data  has  been addressed 
at this   workshop  in  the contribution of K.~Kaneyuki \cite{kaneyuki}. 
This work  analysed  the effects of three 
modifications of  the  neutrino  cross section:
(a) a  change in  $\sigma_{\rm NC}$ of $\pm 30\%$, 
(b) the setting to zero of 
the cross section for    $\pi^\circ$ coherent  production
on a nucleus 
(the process $\nu + A \to \nu + A +\pi^\circ$ \cite{rein1,Baltay}),
and (c) a modification of 
$\pm 30\%$ for the ratio 
 $\sigma_{\rm NQE}/\sigma_{\rm qel}$.
The SK  simulation  was   repeated including   these modifications,
and the  allowed region  in the plane
$(\Delta m^2$,$\sin^2 2 \theta$) was  recalculated  
following the same  standard procedure.
The allowed intervals calculated   with the modified  cross section
are remarkably similar to  the  normal ones.
The effect of the $\pm 30$\% increase 
of the neutral current   cross section corresponds to a small
($\sim \pm 4\%$)  increase in the allowed range for
  $|\Delta m^2|$. This can be understood  noting that
 the neutral current events   contribution to the
single--ring events is  larger for 
$\mu$--like   events, therefore  a larger  neutral current
cross section corresponds to a   larger
$(\mu/e)_{\rm MC}$, to  a smaller  double ratio
$(\mu/e)_{\rm data}/(\mu/e)_{\rm MC}$  and therefore to 
a larger  suppression
of the $\mu$--rate, and corresondingly to a larger $|\Delta m^2|$.  

A $\pm 30\%$  modification of the multi--pion cross
section correspond  to  a shift  of order $\pm 0.02$ in  the minimum  allowed
value of the  mixing parameter 
$\sin^2 2 \theta$. This can be  qualitatively understood
observing that in the presence of a larger contribution of
the non--quasi  elastic  component, the correlation between the 
muon and the neutrino  becomes  weaker.
The Up/Down  asymmetry observed   at the muon level is in  general
smaller than the true asymmetry   at the neutrino  level,  because
of finite  angular  resolution effects,  with the dominant effect being
the  angle
$\theta_{\mu\nu}$    between the
observed muon and the parent neutrino that  is determined  by  the 
interaction cross section.   Assuming a weaker correlation,
the inferred true asymmetry  becomes stronger, and  the mixing parameter
larger.

The effect  of   neglecting the coherent $\pi^\circ$ cross section
is  found  to ne negligible.
These  results  are very encouraging, in the sense that they demonstrate
that  the  atmospheric  neutrino result is  very robust, and  that
the uncertainties in our knowledge of the neutrino cross
section   is    actually very small. This   is  after all not  so surprising
and is a consequence of  the fact that essentially all effects 
in atmospheric  neutrinos  are  visible in ratios
such as $e/\mu$ or up/down, and  the  theoretical uncertainties
tend to  cancel.

A problem  that has  attracted  a  fair  amount 
of attention in the past is the question of the absolute value of the
atmospheric  neutrino fluxes, and the precise shape of  their
energy dependence. Different calculations of the atmospheric
$\nu$ fluxes  \cite{fluka,bartol,honda,honda1,naumov}  yield  results  
  that differ  in absolute  normalization by 
20\% or more,  and also have
non negligible  differences in the  slope of the energy spectrum.
These differences have negligible effects on the
neutrino oscillation  interpretation of the data, however
it would be  very desirable to determine 
the correct normalization of  the fluxes,  but this is  of course
only possible  is the 
absolute value of  $\sigma_\nu$ is known with sufficient  precision.

Future  experiments  with accelerator beams
aiming at more precise  determination of the neutrino oscillation
parameters will very likely have more 
stringent requirements  on  the control of the neutrino cross section.
A quantitative discussion  of this  question is still not complete.
For example, possible effects for the Minos  experiment have been discussed by
Adam Para \cite{para}.
In  experiments that measure the  shape of the 
survival probability $P_{(\nu_\mu \to \nu_\mu)}(E_\nu)$ as a function
of the neutrino energy $E_\nu$  (for a fixed pathlength
$L$)   it is  essential   to  have a good   reconstruction of
 the neutrino energy.
This also  depend to a good understanding of the
neutrino cross section,  to correct   the visible energy for 
the energy  absorbed in   the excitation
of the  target  nucleus, and   the average composition and  multiplicity
of  the final state.

An extremely good  knowledge of the neutrino
cross section will be required   in  case the planned
neutrino factories  will be constructed and used for the
determination of CP violating effects in neutrino flavor 
transitions.  In this case one wants  to compare the 
rates   of processes  such as:
 $\nu_e \to \nu_\mu \to \mu^-$ to 
and
 $\nubar_e \to \nubar_\mu \to \mu^+$ 
and one will  need a control of the 
ratio  $\sigma_\nu / \sigma_{\nubar}$   to a better precision  than the size
of the expected   asymmetries.
An interesting  contribution 
 was given  by Yoshihia Obayashi \cite{obayashi},   for the JHF collaboration.
The JHF  project  \cite{JHF}   plans to use the  50~GeV  proton
beam of the Jaeri  accelerator to  produce an intense
neutrino beam  directed at the Super--Kamiokande  detector
(with a 295~km baseline). The first phase  is  directed at the
determination of $\theta_{13}$  studying 
transitions $\nu_\mu \to \nu_e$, while a second phase
(with a more massive 10~Mt Hyper--Kamiokande detector)
is  aimed  at a study of
CP  violating effects.
A CP violation  asymmetry can be   defined as\footnote{In this
qualitative discussion we will neglect matter effects.}
\begin{equation}
A_{\rm CP} =
{P(\nu_\mu \to \nu_e) - 
P(\nubar_\mu \to \nubar_e) \over
P(\nu_\mu \to \nu_e) + 
P(\nubar_\mu \to \nubar_e) }
\end{equation}
The oscillation probability can  be schematically 
determined as:
\begin{equation}
P(\nu_\mu \to \nu_e) = {N_e \over \sigma_{\nu_e}^{\rm cc} \; \varepsilon_e \;
\Phi_{\mu}^{\rm expected} }
\end{equation}
where
$N_e$ is the observed rate of $e$--like events, 
$\varepsilon_e$ is the detection efficiency and 
 $\Phi_{\mu}^{\rm expected}$ is the expected $\nu_\mu$ fluence
at the detector in the absence of oscillations.
The probability
$P(\nubar_\mu \to \nubar_e)$ can be obtained 
with a  neutrino beam  of  opposite sign.
It is clear that for the determination  of the asymmetry
parameter  only the ratio $\sigma_{\nu_e}/\sigma_{\nubar_e}$  is relevant.
In fact,  considering that also the determination of the expected flux
depends on knowledge of the cross section,  the asymmetry can be rewritten
as a function  of the quantity 
$r_\sigma = 
 (\sigma_{\nu_e}\, \sigma_{\nu_\mu})/
 (\sigma_{\nubar_e}\, \sigma_{\nubar_\mu})$.
A  part of the systematic uncertainties 
cancel in this ratio, however it is not clear
if, even considering this cancellation, the present uncertainties
are sufficiently small to allow the measurement without
a special program of cross section measurements.

\section{Montecarlo Codes}
In all  present and future neutrino  experiments the analysis
of the data will  require detailed   montecarlo  calculations,
therefore the problem of  describing the  neutrino interactions can
be  divided  into two parts of similar importance:
\begin{itemize}
\item The development of a theoretical  framework to 
describe  the interactions.
\item The development of 
montecarlo algorithms to implement  this understanding.
\end{itemize}
Of course, since a calculation of  the neutrino cross sections
with hadrons is not possible from first principles, the first
and most important task is to collect experimentally the necessary
information.
The development of montecarlo algorithms in a non trivial
problem, that has  probably   been given too little 
attention until this moment.
All  the  neutrino experiments  that have 
recently collected data have used   different
montecarlo codes,  written  by members of the
collaboration, and   the results  of the different codes
have not been object of a sufficiently
careful  and detailed comparison.
One of the successes of the NUINT01  workshop, has  been
in fact the  discussion  between different groups and
the comparison between  different  codes.

The Super--Kamiokande  collaboration   has used
two  different montecarlo codes,  the NEUT code \cite{hayato},
originally developed for Kamiokande, and the NUANCE code \cite{casper}
(with a progenitor developed for IMB).
The Soudan--2  detector has used    the code NEUGEN
\cite{ghallagher}  that is  now  developed  to be used by MINOS.
The Icarus  collaboration  has performed  simulations 
using  a  code developed for (and tuned  with the data of)  the Nomad 
detector at CERN \cite{rubbia-mc},  and a second one discussed 
in  \cite{cavanna}.
The Opera  detector  has performed its simulations 
with the code JETTA  developed for
CHORUS,  that  gives  
special attention to the $\nu_\tau$ cross section
\cite{tsukerman},   an additional MC for Opera was presented at the  workshop
\cite{marteau-mc}.  The montecarlo code of Miniboone
was  discussed in \cite{mills}.
This is  not the place for a critical  discussion of the results  of this
comparison,  and only few comments are possible.
No major discrepancy between  the different codes  has been  found,  however
several  non--trivial  effects  have been found, and in fact even 
non  negligible bugs have already been  found and corrected for.
Most codes  tend to use  the same  theoretical  input
namely:  \\
(i)  Llewellyn Smith \cite{Llewellyn} for the
expression  of the quasi--elastic  cross section
on free  nucleons. \\
(ii)  Smith and Moniz for the implementation of  the Fermi   gas model
for quasi--elastic scattering. \\
(iii) The resonance cross section of Rein and Sehgal. \\
(iv) The ``standard'' DIS  formula  for  high $W$ and $Q^2$,
with one of the  publically available sets of PDF's. \\
Even if the theoretical  input  is the same, there are non--trivial
differences that are present,  in particular about  the implementation
of the Fermi--gas model,   the ``joining'' of the
 resonance production and  DIS  scattering regimes,
and the hadronization  of the  final  hadronic  state.
The nuclear  rescattering  effects  are treated 
(when they are) in significantly   different  ways,  so  effects such
as  pion reabsorption,  pion elastic  and  charge--exchange scattering
are very different.
The field  of neutrino  physics  would certainly  benefit
significantly  if some   ``standard''  codes  implementing correctly
 algorithms  recognized as a possible  good
approximations for  the problem,  could become publically available,
as it is the case for  well established  event generators  
in  other  sub--fields of particle physics.
The systematic uncertainties in  the neutrino interaction  properties
are still sufficiently large   that it is probably 
 neither possible nor really desirable to have a  unique ``universal code'',
but in the era of precision studies in neutrino physics,
when the requirements on the event  generators  will be more
stringent, the development of  more  sophisticated 
and better tested instruments  
is  clearly required.

\section{Outlook}
The study of atmospheric neutrinos with energies around
1~GeV has  yielded    remarkable  results  on  the neutrino mass
matrix. Uncertainties on the the neutrino--nucleus cross  section
in this  energy range are  of order $\sim 20\%$. These  uncertainties
very likely have  a small impact on the interpretation of the
data, 
however  for  future studies using accelerator
neutrinos in approximately the same 
range of energy,  a more  precise knowledge of the cross section
would be  valuable, and in fact it will be  required
for  studies that attempt to  detect and measure $CP$ violation effects 
in neutrino  oscillations.
The  size of the  uncertainties on  the $\nu$ interaction properties
can be very likely reduced   incorporating  in our description
of the neutrino cross sections the understanding  of nuclear effects,
hadronic resonance production, and  the behaviour of the 
nucleon structure functions
at low  $Q^2$    that has  been obtained in more recent studies
in electron scattering. 

The near detector  of  the K2K  experiment has collected the
largest existing sample of $\nu$ interactions in the $E_\nu \sim 1$~GeV 
region.
These data  have a great potential to clarify several important
questions on the neutrino interaction properties.
Hopefully a  sufficient precise determination of the  normalization and
energy spectrum  of the $\nu$--beam will allow   the measurement of
absolute values for the cross sections;
in  any case  important 
information about  the  ratios of cross sections
can be obtained. 
The measurement of the 
rate  for the neutral  process $\nu + A \to \nu + A + \pi^\circ$ 
relative (for example) to quasi--elastic scattering, is very
important to   put limits on oscillations 
of standard neutrinos into sterile states.

Some natural goals  for future studies  could be:
\begin{itemize}
\item A  more  quantitative  estimate of the importance
of uncertainties on $\sigma_\nu$  as limiting factor
for  the sensitivity 
of future experiments  on neutrino oscillations.

\item A complete analysis of the K2K near detector data to
obtain the most of these  remarkable results.

\item  The development of a  treatment of nuclear effects
that incorporates   modern  understanding and  calculation methods,
going beyond the Fermi gas approximation.

\item  More accurate  modeling
of pion and nucleon rescattering in nuclear matter.

\item  An improved treatment of  hadronic  resonance production,
with a  smooth joining  of this kinematical region to the
deep inelastic region, using the insight allowed by the electron--scattering
results.

\item Development  and testing of  the montecarlo  algorithms
that implement our knowledge of the  $\nu$ interaction properties.

\item  Last but not least:  
the proposal of new  experimental studies aimed
at the measurement of the most important/interesting   properties
of neutrino interactions, using the new  intense $\nu$--beams.
See for example the contribution of Jorge Morfin in these Proceedings
\cite{morfin}.
\end{itemize}

Neutrinos can be a precious 
window on  the physics of the unification (beyond the standard model),
but on the other  hand they can also
be an extraordinarily valuable probe on  the properties of
QCD  and  the structure 
of the nucleons. The $(V-A)$ structure of the  neutrino
interactions  allows to explore   with an axial probe the properties of 
a  bound system of  quarks in ways that are   very difficult
to obtain  in  $e$--scattering. The calculation  of the 
nucleon structure functions and their evolution  
in the low $Q^2$  region  remains an important  challenge for
future QCD studies.
In summary  the measurement of the  $\nu$ and $\nubar$ interaction
properties can be seen  both as an instrument necessary to  determine
with high precision the $\nu$ masses and mixing {\em and} as a 
a important  scientific goal per se.
In the golden era of  neutrino physics, the motivations
are powerful,  the difficulties challenging  and the
opportunities  rich e varied.

\vspace{0.2 cm}
\noindent {\bf Acknowledgments}
I'm very grateful  to the organizers of NUINT01  for a very
interesting  and  frutiful
workshop, and to many of the participants for
instructive discussions.
Special thanks  to
Maurizio Lusignoli
Makoto Sakuda,
and Chris Walter.


\begin{thebibliography}{99}
\bibitem{ichikawa}
A.~Ichikawa,
``New near detector of
K2K, its physics goals
and potential.",
these Proceedings.
\bibitem{wood1}
S.~Wood,
``Experimentation at JLAB:
what are the existing
facilities/experiments,
potential for            
modifications and
upgrades, potential for
new initiatives",
these Proceedings.
\end{thebibliography}
\end{document}